# Search for magnetoelectric monopole response in $Cr_2O_3$ powder

Syed Q. A. Shah, Ather Mahmood, Arun Parthasarathy, and Christian Binek[*]

*Department of Physics & Astronomy and the Nebraska Center for Materials and Nanoscience, University of Nebraska-Lincoln, Lincoln, NE 68588-0299, USA*

*To whom correspondence should be addressed. E-mail: cbinek@unl.edu

**Abstract:**

Powder samples have been suggested as a pathway to fabricate isotropic magnetoelectric (ME) materials which effectively only have a pseudoscalar or monopole ME response. We demonstrate that random distribution of ME grains alone does not warrant isotropic ME response because the activation of a non-vanishing ME response requires a ME field cooling protocol which tends to induce preferred axes. We investigate the evolution of ME susceptibility in powder chromia samples for various ME field cooling protocols both theoretically and experimentally. In particular, we work out the theoretical expressions for ME susceptibility for powder Chromia in the framework of statistical mechanics where Boltzmann factors weigh the orientation of the Néel vector relative to the local orientation of the *c*-axis of a grain. Previous approximations oversimplified the thermodynamic nature of the annealing process giving rise to misleading conclusions on the role of the magnitude of the applied product of electric and magnetic fields on the ME response. In accordance with our refined theory, a strong dependence of the functional form of α vs. *T* of Chromia powders on the ME field cooling protocol is observed. It shows that Chromia powder is not generically an isotropic ME effective medium but provides a pathway to realize the elusive isotropic ME response.

**Introduction**:

The magnetoelectric (ME) effect was first proposed by Landau and Lifshitz in 1958 [1]. In a linear ME medium, where time and spatial inversion symmetry are broken while their combined action leaves the system invariant, the magnetic $\mathbf{H}$-field induces an electric polarization $\mathbf{P}$ according to $\mathbf{P_i} = \alpha_{ij}\mathbf{H_j}$. Conversely, magnetization, $\mathbf{M}$, is induced by an applied electric $\mathbf{E}$-field according to $\mu_0\mathbf{M_i} = \alpha_{ji}\mathbf{E_j}$ with $\alpha_{ij}$ being the ME tensor. The linear ME antiferromagnetic $Cr_2O_3$ has a rhombohedral unit cell and a magnetic point group $\overline{3}'m'$. This symmetry allows for a diagonal ME tensor with ME coupling terms $\alpha_\parallel$ and $\alpha_\perp$ for responses parallel and perpendicular to its trigonal crystal axis [2]. The $Cr_2O_3$ ME susceptibility tensor, $\alpha$, can be decomposed into a trace free and a pseudoscalar component according to:

$$\alpha = \begin{pmatrix} \alpha_\perp & 0 & 0 \\ 0 & \alpha_\perp & 0 \\ 0 & 0 & \alpha_\parallel \end{pmatrix} = \frac{1}{3}(\alpha_\perp - \alpha_\parallel)\begin{pmatrix} 1 & 0 & 0 \\ 0 & 1 & 0 \\ 0 & 0 & -2 \end{pmatrix} + \theta\begin{pmatrix} 1 & 0 & 0 \\ 0 & 1 & 0 \\ 0 & 0 & 1 \end{pmatrix} \qquad (1)$$

Here $\theta = \frac{1}{3}(2\alpha_\perp + \alpha_\parallel)$ is a non-zero pseudoscalar component for $Cr_2O_3$ [3-6]. In a modern view on magnetoelectricity, the ME susceptibility tensor is often considered in terms of multipolization. In the context of multipolization, the trace free tensor component is known as quadrupole and the pseudoscalar or axion piece is known as monopole contribution [7]. The possibility of a material with a non-zero axion piece was long debated mainly due to the notion of a Post constraint [8], but later unambiguously confirmed to be present in $Cr_2O_3$ [9]. Once the presence of a non-zero axion piece was established, the search for a cubic ME crystal began where $\alpha_\perp = \alpha_\parallel$. The isotropic ME response in a single crystal remained elusive. As an alternative it has been suggested to use powders of ME single crystals with diagonal ME response, to create an effective medium with isotropic ME properties. Shtrikman and Treves pointed out that for a powder of a diagonal ME material with $\frac{\alpha_\parallel}{\alpha_\perp} > 0$ near the Néel temperature, $T_N$, the ME response will be isotropic when ME annealing the powder in collinear applied electric and magnetic fields [10]. This led Veremchuck et al. to the conclusion that the ME effect in chromia powder samples is isotropic [11]. However, neither is $\frac{\alpha_\parallel}{\alpha_\perp} > 0$ fulfilled in chromia or any known magnetoelectric nor is the approach by Shtrikman and Treves a true thermodynamic theory which utilizes thermal averaging. It thus fails to describe the proper dependence of the ME response of a powder on the magnitude of the applied fields during the ME annealing process. We show in this work that such a strong dependence exists and that chromia powder is not a priory an isotropic ME effective medium.

More recently, isotropic ME response gained new attention in the context of axion electrodynamics and the topological ME effect which can be realized in topological insulator (TI) materials [12]. Hence, it is desirable to establish a deeper connection between axion dynamics in TIs and in topologically trivial ME insulators with the central question being whether it is possible to find a topologically trivial material which has only an axion piece ME response [13-15]. It is worth noting that the isotropic ME response in TIs with realistic disorder decays quickly over time. In addition, the magnetic ordering temperature for magnetic TIs is very low ~ 15K [16, 17] such as Cr-doped $(Bi, Sb)_2Te_3$, making the investigation of the corresponding monopole response near room temperature in topologically trivial insulators even more attractive [18]. The unifying aspect of isotropic ME response in all material classes is the possibility to map the peculiar property of the constitutive equation onto a modified form of the Maxwell action. Maxwell's field equations in electromagnetism are the Euler-Lagrange equations of the Maxwell action. For the case of isotropic ME materials, the Lagrangian density contains an additional term, $\theta \mathbf{E} \cdot \mathbf{B}$, which couples the $\mathbf{E}$ and $\mathbf{B}$ fields, via the axion field $\theta$. The Maxwell action of such ME condensed matter systems is analogous to the action of axion field models in particle physics [19, 20].

The archetypal ME antiferromagnet $Cr_2O_3$ has a diagonal ME tensor. Hence, powder of $Cr_2O_3$ with randomly distributed orientation of the $c$-axis of individual grains is an ideal candidate to look for near room temperature isotropic ME response. There is a lot of work reported in literature for textured thin films and $Cr_2O_3$ single crystal [21-25] but very little for the polycrystalline $Cr_2O_3$ powder. In 1962 Shtrikman *et al.* observed the ME effect in highly compressed polycrystalline $Cr_2O_3$ powders [10]. Very few subsequent studies show that the hot pressed polycrystalline $Cr_2O_3$ powder gives ME response that is one-third of the $Cr_2O_3$ single crystal [26, 27]. Shtrikman *et al.* also calculated the theoretical expressions of the ME susceptibility, $\alpha$, for both parallel and perpendicular relative orientation of the applied electric and magnetic annealing fields. However, their theory of ME annealing was based on the approximation that all grains will align their Néel vector in a way which minimizes their energy leaving out the thermodynamic aspect that excited states have a non-zero probability to be populated determined by Boltzmann factors. In our work we experimentally investigate the evolution of the ME susceptibility, α, in polycrystalline $Cr_2O_3$ powders with different annealing field protocols and compare the results with an improved thermodynamic model of the ME susceptibilities. By doing so, we show that ME annealing affects the functional form of the temperature dependence of the ME response which shows a pathway towards the realization of an isotropic ME effective medium.

**Experiment:**

We used a highly compressed micron size polycrystalline $Cr_2O_3$ power sample which has a density 5.21 g/cm$^3$ comparable to that of single crystal *i.e*., 5.22 g/cm$^3$. The sample was cut from the $Cr_2O_3$ sputtering target (Kurt. J. Lesker company part number EJTCROX282A4) in the shape of a cuboid with dimensions of 5mm × 5mm × 2.5mm. The purpose of using micron size grains rather than submicron grains was to minimize parasitic ferromagnetic contributions from uncompensated spins at the surfaces of crystallites, reported in the literature [10]. These parasitic moments can be much larger than the ME induced moment and thus contribute to background and noise in characterizing the ME effect.

A Quantum Design, Superconducting Quantum Interface Device (SQUID) MPMS-XL was used as a platform to apply **H**-fields (range 0 to 5.57× 10$^6$ A/m), provide temperature control and detection of the ME effect. An external direct current (DC) power supply was used to apply the **E**-field simultaneously with **H**-field while annealing the sample from 370 K ($T$>>$T_N$) to 20 K ($T$<<$T_N$). The SQUID was used to measure the ME response of the polycrystalline $Cr_2O_3$ sample with the help of a self-implemented alternating current (AC) AC-SQUID technique [28]. The AC technique has the advantage of being insensitive to any static magnetization background from uncompensated moments which, in a DC measurement, can mask the small ME signal. Our AC technique uses an external AC power supply to apply a low frequency (1Hz) AC voltage across the electrodes of the sample which is kept stationary within the center of the gradiometer pickup coil of the SQUID [29]. The AC electric field excitation is applied to the sample via copper wires which are attached to the surface and the bottom of the sample cuboid with the help of silver paste electrodes. The ME response of the sample on the alternating applied **E**-field generates alternating magnetic flux which induces an AC voltage in the SQUID pickup coils. The alternating output signal from the rear panel of MPMS-XL is fed into a lock-in amplifier (model SR830). The reference signal is provided by the AC power supply. Rather than measuring the signal of the x-channel after establishing the phase relation which maximizes the signal, we use the lock-in amplifier in polar mode $(R, \Theta)$ with $(x = R\ Cos\ \Theta, y = R\ Sin\ \Theta)$. Measuring $R$ makes the output independent of the relative phase between reference and input signal. The reason to do this is that after each parameter change, such as changes of the sample temperature, the phase between signal and reference changes unpredictably due to the automatic reset the commercial SQUID sensor performs. At low frequencies, *i.e.*, in the virtual absence of losses, $R$ is an accurate measure for the ME susceptibility with the only caveat being that temperature dependent changes of the sign of the susceptibility reveal themselves in $R$ as reflection from the temperature axis rather than crossing of the temperature axis. The first harmonic of the $R$-signal is a linear function of the applied AC voltage amplitude in accordance with the linear ME effect. To further improve the signal-to-noise ratio we measure at each temperature the ME response for the five AC-voltage amplitudes of 70, 140, 210, 280, and 350 V. We code a LabVIEW program which records the average of twenty different readings of SQUID voltage from the lock-in amplifier in response of a particular AC-voltage applied across the sample. The magnetic response of the sample in terms of voltage, read from lock-in amplifier, is plotted as a function of E-field corresponding to five different AC-voltage amplitudes. From the relation $\mu_0 \boldsymbol{M} = \alpha \boldsymbol{E}$ the slope of a linear best fit to the five isothermal data points provides the ME susceptibility at a given temperature. In order to obtain a specific result which enables comparison of different samples we normalize the lock-in signal with respect to the mass of the respective sample. The schematic diagram of experimental setup is shown in **Fig. 1**.

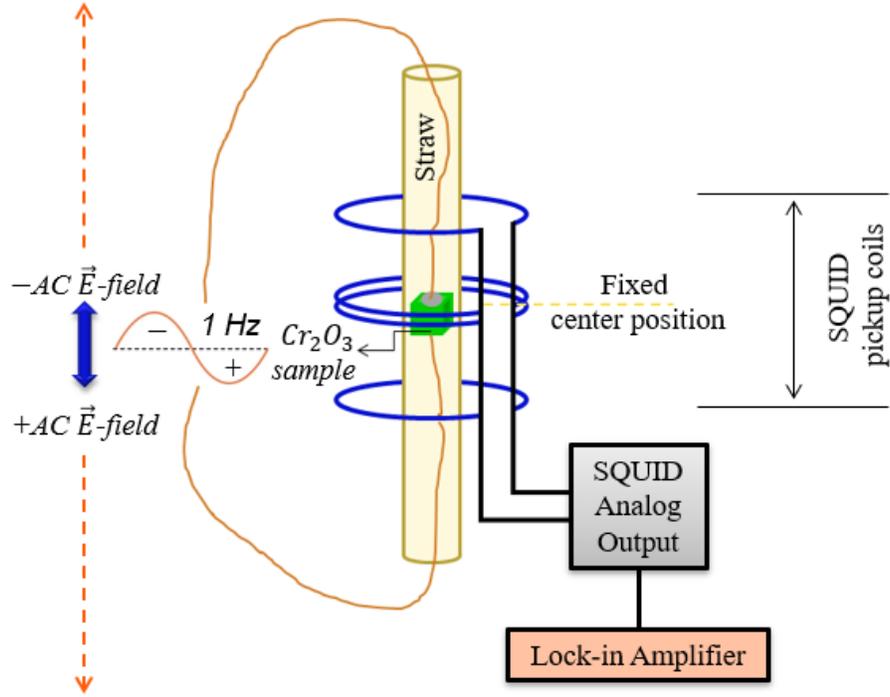

**Figure 1:** Schematic diagram of custom build AC-SQUID MPMS-XL.

In polycrystalline $Cr_2O_3$ powder, the crystal axes of the individual grains are randomly oriented in space with rotational symmetry. The ME susceptibility tensor for the polycrystalline powder can be characterized by a nine-component tensor where $\alpha_{ij} = \alpha_{ji}$.

$$\alpha = \begin{pmatrix} \alpha_{xx} & \alpha_{xy} & \alpha_{xz} \\ \alpha_{yx} & \alpha_{yy} & \alpha_{yz} \\ \alpha_{zx} & \alpha_{zy} & \alpha_{zz} \end{pmatrix} \qquad (2)$$

When cooling the $Cr_2O_3$ powder sample in zero applied fields from above to below the Néel temperature, $T_N$=307 K, the powder will have equal distribution of $180^{\circ}$ antiferromagnetic domains states, which means for each grain where the Néel vector points along the $c$-axis in positive direction there is a grain where the Néel vector points in the opposite direction. This gives rise to a net zero ME response. However, when **E** and **H** fields are simultaneously applied during the cooling process from $T>>T_N$ to $T<<T_N$, the degeneracy of the domains is lifted, and a temperature dependent ME response can be observed below 307 K. Activation of tensor components depends on the relative orientation of applied **E** and **H** fields while cooling. Cooling of the polycrystalline $Cr_2O_3$ powder under the application of parallel **E** and **H** fields will only activate the diagonal components of ME susceptibility tensor (2) **[10]**. Isotropic or pure pseudoscalar response is a special case. As a necessary condition it requires parallel field annealing to eliminate off-diagonal elements in Eq. (2). However, it requires $\alpha_{xx} = \alpha_{yy} = \alpha_{zz}$ in addition which is without further measures not the case. The off-diagonal components $\alpha_{ij}$ of the ME tensor (2) are activated when the polycrystalline $Cr_2O_3$ powder is cooled under the application of simultaneously applied perpendicular **E** and **H** fields with $\boldsymbol{E} = E_0 \, \vec{e}_i$ and $\boldsymbol{H} = H_0 \, \vec{e}_j$.

Measuring the ME susceptibility response for the diagonal component $\alpha_{zz}$ requires two steps. First the powder is ME annealed, *i.e.*, cooled through $T_N$ in the presence of simultaneously applied parallel **E** and **H** fields both along $\hat{\boldsymbol{z}} -$ axis. The $\hat{\boldsymbol{z}} -$ axis is defined by the cylinder axis of the gradiometer pickup coils shown in Fig. 1. This annealing protocol will only leave the diagonal components of the ME tensor (2) to be non-zero with $\alpha_{xx} = \alpha_{yy} \neq \alpha_{zz}$. In a second step, the ME response $\alpha_{zz}$ is measured by applying a low frequency (~1Hz ) AC-voltage along the $\hat{\boldsymbol{z}} -$ axis. For improved signal-to noise ratios the AC-voltage is applied with various amplitudes ranging between 70 V – 350 V with a step size of 70 V. In the low frequency regime ME response and AC excitation are in phase and the first harmonic lock-in signal normalized to the mass of the respective sample is proportional to $< \alpha_{zz} >$ where $<...>$ indicates that the integral susceptibility measurement averages over the distribution of random grain orientations of the sample. Because we perform the lock-in detection in polar coordinates to avoid the need for a phase alignment, we plot $|< \alpha_{zz} >| vs \, T$ in **Fig. 2**. The zero crossing of $< \alpha_{zz} > vs \, T$ manifests as touching of the temperature axis of the $|< \alpha_{zz} >| vs \, T$ data. The zero-crossings temperatures for different **E.H** products are shown in the inset of **Fig. 2**. There is a clear dependence of the zero-crossing temperature on the magnitude of the annealing field products. The dependence of the zero-crossing temperatures on the annealing field products is very significant because it shows that the qualitative $T$-dependence of $< \alpha_{zz} >$ can be affected by the annealing protocol. This in turn hints at the possibility to achieve isotropic ME response. We will show in our more in-depth analysis below that a zero-crossing at around 168 K would indicate isotropic response.

**F**igure 3 shows the experimental findings of the temperature dependent response for the off-diagonal component, $< \alpha_{zx} > \, vs \, T$. Again, the ME susceptibility response for the off-diagonal component $< \alpha_{zx} >$ can be obtained in two steps, first by cooling the powder through $T_N$ in the presence of simultaneously applied perpendicular **E** (along $\hat{x} -$ axis) and **H** (along $\hat{z} -$ axis) fields. The ME response for $< \alpha_{zx} >$ is then obtained by applying a low frequency ~1Hz AC-voltage ranging $70V - 350$ V with a step size of 70 V along the $\hat{x} -$ axis. There is a clear increase in the magnitude of the ME susceptibilities with an increase in the annealing field product for both $< \alpha_{zz} >$ and $< \alpha_{zx} >$. This evolution of ME susceptibility components for polycrystalline $Cr_2O_3$ powder with annealing field products has been completely ignored in the previous study **[10]** and, to the best of our knowledge, has not been reported in the literature before. Below we work out the refined theoretical expressions for the ME susceptibility components for polycrystalline $Cr_2O_3$ powder by incorporating the Boltzmann statistics.

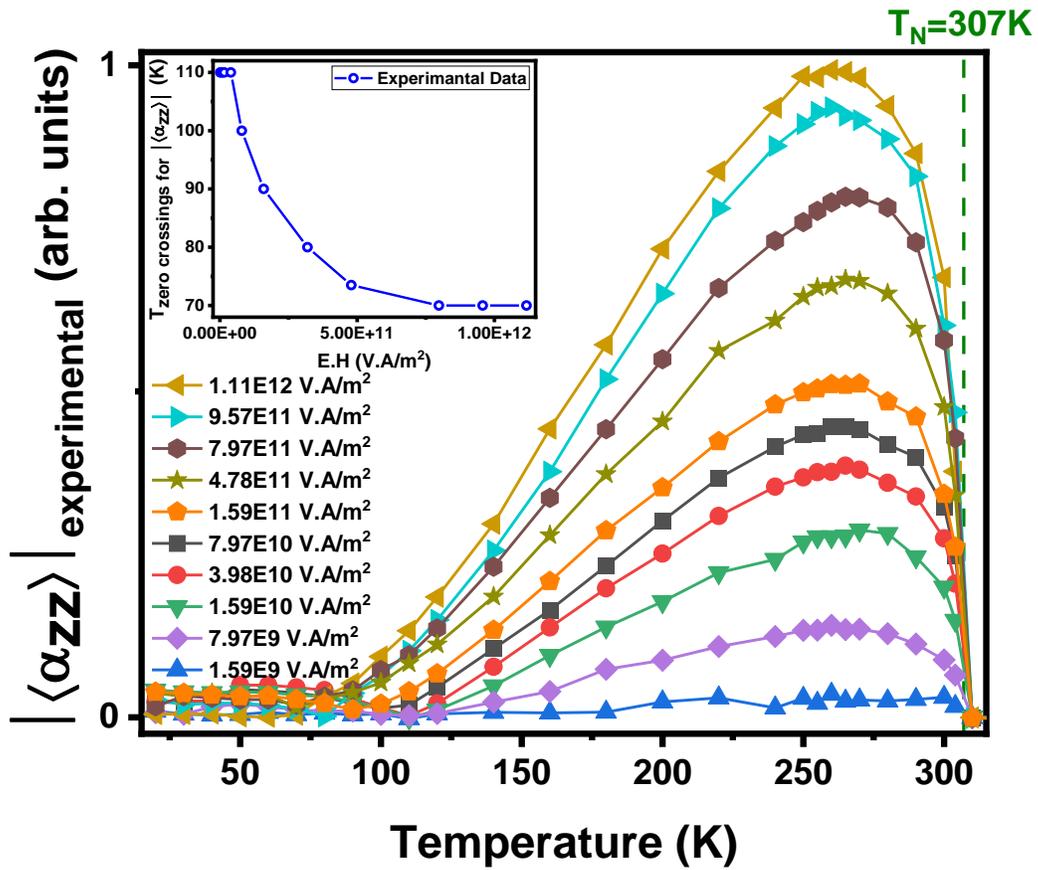

**Figure 2:** Measured $|< \alpha_{zz} >|$ vs T for various **E.H** products. The **E**-field is constant at $2\times10^5$ V/m whereas **H**-fields change from H= $7.95\times10^3$ A/m to H= $5.57\times10^6$ A/m giving rise to **E.H** products $1.59 \times 10^9 \text{VA/m}^2 \leq EH \leq 1.11 \times 10^{12} \text{VA/m}^2$ . The inset shows the temperature dependence of the zero-crossings of $|< \alpha_{zz} >|$ for various **E.H** products.

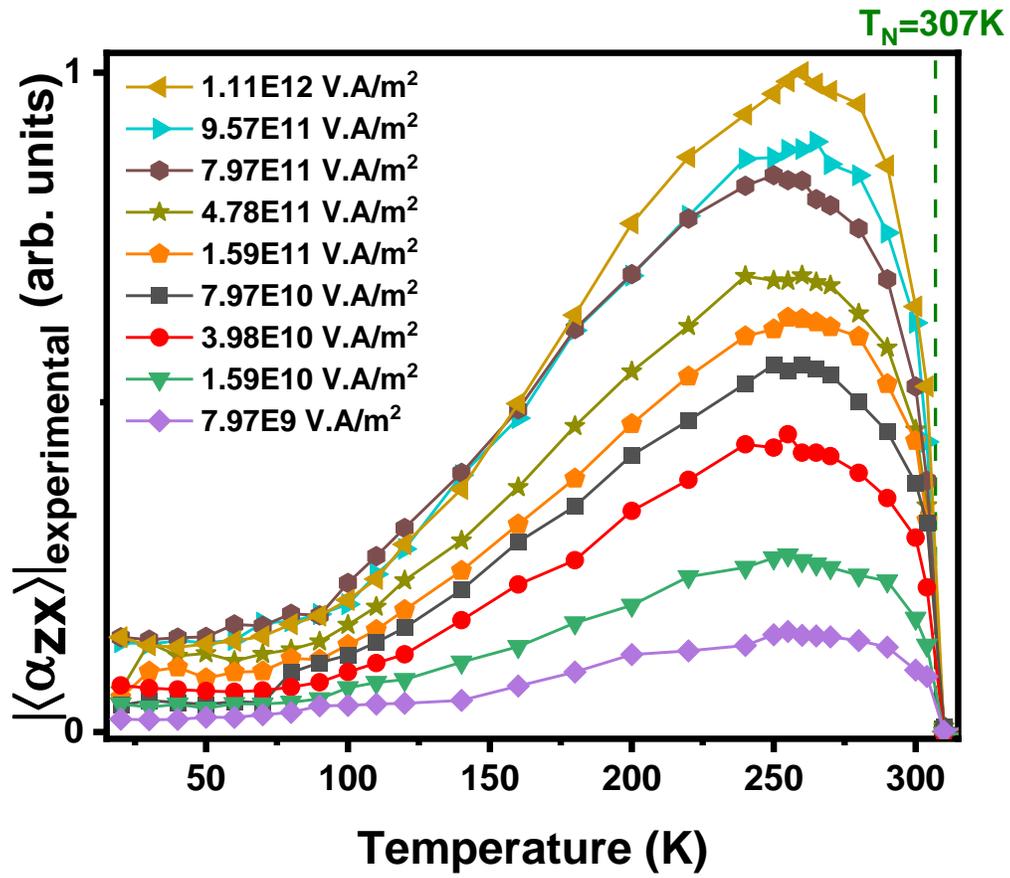

**Figure 3:** Measured $< \alpha_{zx} >$ vs T for various **E.H** products. **H**-fields change from H= $7.95 \times 10^3$ A/m to H= $5.57 \times 10^6$ A/m giving rise to **E.H** products $1.59 \times 10^9 \text{VA/m}^2 \leq EH \leq 1.11 \times 10^{12} \text{VA/m}^2$

**Thermodynamics of annealing and distribution average of ME response of Cr₂O₃ powder**

A single crystal of Cr₂O₃ can be in two degenerate single domain states. In a single domain state, the crystal is antiferromagnetically long range ordered with collinear alignment of the spins of the 4 sublattices along the magnetic easy axis which is the *c*-axis of the rhombohedral crystal. The two degenerate states are, hereafter called $(+)$ and $(-)$ states. They are distinguished by opposite orientation of the corresponding Néel vector. The $180^0$ single-domain states $(+)$ and $(-)$ of Cr₂O₃ have respective ME response $\alpha^+$ and $\alpha^- = -\alpha^+$, each defined by two independent coordinates for directions parallel $\parallel$ and perpendicular $\perp$ to the trigonal crystal axis.

$$\begin{cases} (+): (\alpha_\parallel, \alpha_\perp) \\ (-): (-\alpha_\parallel, -\alpha_\perp) \end{cases} where\ \alpha_\perp < 0 \qquad (3)$$

In polycrystalline powder, the crystal axes of the grains are randomly oriented in space making the material invariant under rotation. The local coordinate system of a grain whose crystal axis is oriented at $(\theta, \phi)$ as shown in **Fig. 4** is defined by the orthonormal basis.

$$\widetilde{x} = \begin{pmatrix} Cos\theta Cos\phi \\ Cos\theta Sin\phi \\ -Sin\theta \end{pmatrix}, \ \widetilde{y} = \begin{pmatrix} -Sin\phi \\ Cos\phi \\ 0 \end{pmatrix}, \ \widetilde{z} = \begin{pmatrix} Sin\theta Cos\phi \\ Sin\theta Sin\phi \\ Cos\theta \end{pmatrix} \qquad (4)$$

where the unit vector $\widetilde{z}$ is pointing along the direction of the crystalline *c*-axis and vectors $\widetilde{x}$, and $\widetilde{y}$ are in the plane perpendicular to the crystal axis. The $\hat{z}$ − axis of the reference coordinate system is the axis of the gradiometer pickup coils. The $\hat{x}$ and $\hat{y}$ − axis are arbitrary but fixed axes perpendicular to $\hat{z}$ − axis and each other.

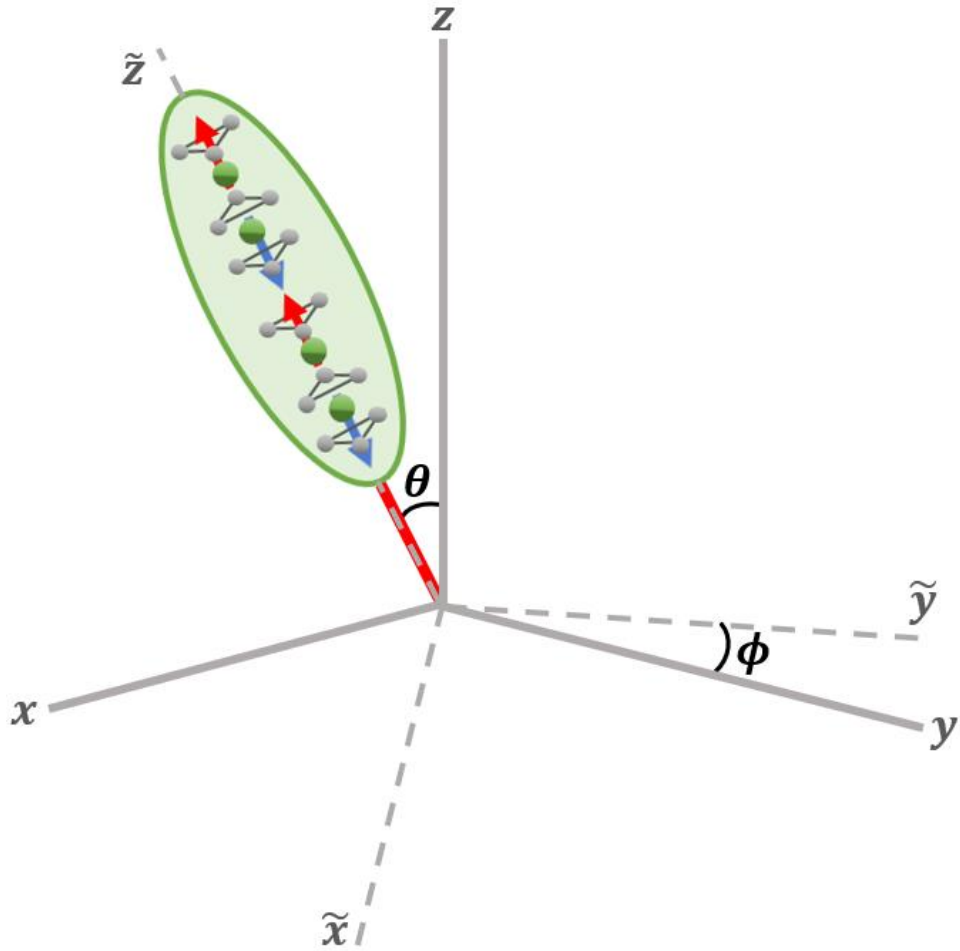

**Figure 4:** Local coordinate system of a grain of polycrystalline $Cr_2O_3$ powder whose crystal axis is oriented at $(\theta, \phi)$.

The free energy of the grain allows for two ME coupling terms for applied **E** and **H** fields:

$$\mathcal{F}_{ME}^{\pm}(\theta, \phi) = \mp[\alpha_{\perp}(\boldsymbol{E}.\widetilde{\boldsymbol{x}})(\boldsymbol{H}.\widetilde{\boldsymbol{x}}) + \alpha_{\perp}(\boldsymbol{E}.\widetilde{\boldsymbol{y}})(\boldsymbol{H}.\widetilde{\boldsymbol{y}}) + \alpha_{\parallel}(\boldsymbol{E}.\widetilde{\boldsymbol{z}})(\boldsymbol{H}.\widetilde{\boldsymbol{z}})] \tag{5}$$

In general, we can write the ME susceptibility in terms of $\mathcal{F}_{ME}$:

$$\alpha_{ij}^{\pm}(\theta, \phi) = {}^{-\mathcal{F}_{ME}^{\pm}}\!\big/\!{}_{(E_i H_j)} \tag{6}$$

where the diagonal components for (+) domain are:

$$\begin{cases} \alpha_{xx}^{+} = (\alpha_{\parallel}Sin^2\theta + \alpha_{\perp}Cos^2\theta)Cos^2\phi + \alpha_{\perp}Sin^2\phi \\ \alpha_{yy}^{+} = (\alpha_{\parallel}Sin^2\theta + \alpha_{\perp}Cos^2\theta)Sin^2\phi + \alpha_{\perp}Cos^2\phi \\ \alpha_{zz}^{+} = \alpha_{\parallel}Cos^2\theta + \alpha_{\perp}Sin^2\theta \end{cases} \tag{7}$$

And the off-diagonal components for (+) domain are:

$$\begin{cases} \alpha_{xy}^{+} = \alpha_{yx}^{+} = (\alpha_{\parallel} - \alpha_{\perp})Sin^2\theta Sin\phi Cos\phi \\ \alpha_{yz}^{+} = \alpha_{zy}^{+} = (\alpha_{\parallel} - \alpha_{\perp})Sin\theta Cos\theta Sin\phi \\ \alpha_{zx}^{+} = \alpha_{xz}^{+} = (\alpha_{\parallel} - \alpha_{\perp})Sin\theta Cos\theta Cos\phi \end{cases} \tag{8}$$

Let "$n$" be the total number of grains within the sample, among which $n^{+}$are in (+) domain state and $n^{-}$ are in ($-$) domain state such that.

$$n = n^{+} + n^{-} \tag{9}$$

Also assume among those "$n$" grains $\delta n^{+}(\theta, \phi)$ and $\delta n^{-}(\theta, \phi)$ are infinitesimal small number of grains distributed among the two (+) and (–) domains, with $c$-axes within a small solid angle $d\Omega$ around (θ, ϕ), where $d\Omega = Sin\theta d\theta d\phi$. The value for $\theta$ goes from $0 \to \pi/2$ and $\phi$ goes from $0 \to 2\pi$. One can express $\delta n^{+}(\theta, \phi)$ and $\delta n^{-}(\theta, \phi)$ in terms of $n^{+}$ and $n^{-}$ as

$$\delta n^{\pm}(\theta, \phi) = n^{\pm} \, d\Omega \tag{10}$$

The probability of finding a grain in any of the particular domain state either (+) or ($-$) within $d\Omega$ is

$$p^{\pm}(\theta, \phi) = \frac{\delta n^{\pm}(\theta, \phi)}{\delta n^{+}(\theta, \phi) + \delta n^{-}(\theta, \phi)} \tag{11}$$

with sum of probability equals to unity i.e.,

$$p^{+}(\theta, \phi) + p^{-}(\theta, \phi) = 1 \tag{12}$$

Substituting equation (10) in (11) results

$$p^{\pm}(\theta, \phi) = \frac{n^{\pm}}{n^{\pm} + n^{\pm}} \tag{13}$$

If $\alpha_{ij}^{\pm}$ is the ME susceptibility response of the $\delta n^{\pm}(\theta, \phi)$ grains which lies within $d\Omega$ then the total ME susceptibility response of all the grains within the solid angle $d\Omega$ i.e., $\alpha_{ij}(\theta, \phi)$, is

$$\alpha_{ij}(\theta, \phi) = \alpha_{ij}^{+} p^{+}(\theta, \phi) + \alpha_{ij}^{-} p^{-}(\theta, \phi) \tag{14}$$

Here $i$ & $j$ are the directions in which ME susceptibility is measured. Using the condition $\alpha_{ij}^{+} + \alpha_{ij}^{-} = 0$, equation (14) becomes

$$\alpha_{ij}(\theta, \phi) = \alpha_{ij}^{+} \left( p^{+}(\theta, \phi) - p^{-}(\theta, \phi) \right) \tag{15}$$

Substituting $p^{+}(\theta, \phi)$ and $p^{-}(\theta, \phi)$ from equation (13), the equation (15) can be rewritten as

$$\alpha_{ij}(\theta, \phi) = \alpha_{ij}^{+} \left( \frac{n^{+}}{n^{+} + n^{-}} - \frac{n^{-}}{n^{+} + n^{-}} \right) = \alpha_{ij}^{+} \left( \frac{n^{+} - n^{-}}{n^{+} + n^{-}} \right) \tag{16}$$

According to the Boltzmann statistics $n^{-}$ can be defined in terms of $n^{+}$ as

$$n^{-} = n^{+} e^{-\frac{2\alpha_{kl}^{+} E_k H_l}{k_B T_N}} \tag{17}$$

Here $k$ & $l$ are the directions in which **E** and **H** fields are applied during annealing the powder.

Using equation (17) the equation (16) can be expressed as

$$\alpha_{ij}(\theta,\phi) = \alpha_{ij}^+(T)Tanh\left[\frac{\alpha_{kl}^+(T_N^-)E_kH_lV}{k_BT_N}\right] \tag{18}$$

The fractions of + and – domains are determined during field cooling through $T_N$, where $T_N^-$ in (18) is the temperature just below the $T_N$. At $T_N^-$ the domain state of a grain is selected and freezes on cooling near $T_N$. The average ME susceptibility response of the powder can be calculated by the volume average.

$$\langle\alpha_{ij}\rangle = \frac{1}{\frac{4\pi R^3}{3}}\int_0^R r^2 dr \int_0^{2\pi}\int_0^\pi \alpha_{ij}(\theta,\phi)Sin\theta d\theta d\phi = \frac{1}{4\pi}\int_0^{2\pi}\int_0^\pi \alpha_{ij}(\theta,\phi)Sin\theta d\theta d\phi \tag{19}$$

**Theoretical expression for the average diagonal component $<\alpha_{zz}>$:**

The ME susceptibility response $\alpha_{zz}$ of a single grain can be calculated by using equation (18) for: $i = j = z$ and $k = l = z$ which yields.

$$\alpha_{zz}(\theta,\phi) = \alpha_{zz}^+(T)Tanh\left[\frac{\alpha_{zz}^+(T_N^-)E_zH_zV}{k_BT_N}\right] \tag{20}$$

Using the expression for $\alpha_{zz}^+$ from equation (7) and substituting in equation (20) yields

$$\alpha_{zz}(\theta) = \{\alpha_\parallel(T)Cos^2\theta + \alpha_\perp Sin^2\theta\}Tanh\left[\frac{(\alpha_\parallel(T_N^-)Cos^2\theta + \alpha_\perp(T_N^-)Sin^2\theta)E_zH_zV}{k_BT_N}\right] \tag{21}$$

The average response of the powder can be calculated by using equation (19)

$$\langle\alpha_{zz}(T)\rangle = \frac{1}{2}\int_0^\pi \alpha_{zz}(\theta)Sin\theta d\theta = \int_0^{\pi/2}\alpha_{zz}(\theta)Sin\theta d\theta$$

The expression

$$Tan^2\vartheta = -\frac{\alpha_\parallel(T_N^-)}{\alpha_\perp(T_N^-)} \tag{22}$$

signifies the angle where the thermodynamic ground state at $T=T_N^-$ switches from $\alpha_\parallel > 0 \; and \; \alpha_\perp < 0$ to $\alpha_\parallel < 0 \; and \; \alpha_\perp > 0$ [10]. Splitting the integral around this angle $\vartheta$ yields

$$\langle \alpha_{zz}(T) \rangle = \int_0^\vartheta \alpha_{zz}(\theta) Sin\theta d\theta + \int_\vartheta^{\pi/2} \alpha_{zz}(\theta) Sin\theta d\theta \qquad (23)$$

Substituting $\alpha_{zz}(\theta)$ from equation (21) and making the approximation that the sign change of the ME response at $\vartheta$ is applicable for $T > 0$ yields

$$\langle \alpha_{zz}(T) \rangle = \int_0^\vartheta Tanh\left[\left\{\frac{\alpha_\parallel(T_N^-)+\alpha_\perp(T_N^-)}{2} + \frac{\alpha_\parallel(T_N^-)-\alpha_\perp(T_N^-)}{2}Cos2\theta\right\}\frac{E_z H_z V}{k_B T_N}\right]\left\{\left(\alpha_\parallel(T) - \alpha_\perp(T)\right)Cos^2\theta + \alpha_\perp\right\}Sin\theta d\theta - \int_\vartheta^{\pi/2} Tanh\left[\left\{\frac{\alpha_\parallel(T_N^-)+\alpha_\perp(T_N^-)}{2} + \frac{\alpha_\parallel(T_N^-)-\alpha_\perp(T_N^-)}{2}Cos2\theta\right\}\frac{E_z H_z V}{k_B T_N}\right]\left\{\left(\alpha_\parallel(T) - \alpha_\perp(T)\right)Cos^2\theta + \alpha_\perp\right\}Sin\theta d\theta$$

According to the Cauchy mean value theorem there exists a unique value of $\theta$ where the hyperbolic tangent can be moved outside of the integrals. The Tanh-term introduces temperature-dependent scaling factors $C_1$ and $C_2$. From experimental data of the temperature dependence of the parallel and perpendicular ME susceptibility we know that for temperatures near $T_N$, $\alpha_\parallel(T_N^-) \gg -\alpha_\perp(T_N^-)$ [30]. From condition (22) we can conclude that this implies $\vartheta$ to be only slightly smaller than $\pi/2$.

With this we obtain $C_1 \approx Tanh\left[\frac{\alpha_\parallel(T_N^-)E_z H_z V}{2k_B T_N}\right]$ and $C_2 \approx -Tanh\left[\frac{\alpha_\perp(T_N^-)E_z H_z V}{k_B T_N}\right]$ when using the approximations that for the integral between $[0,\vartheta]$ the Cauchy mean value theorem leads to the approximation $Cos2\theta' \approx 0$ for $\theta' \approx \frac{\vartheta}{2} \approx \frac{\pi}{4}$ and for the integral between $[\vartheta,\pi/2]$ one finds $Cos2\theta'' \approx -1$ for $\theta'' \approx \pi/2$. Together with $\alpha_\parallel(T_N^-) \gg -\alpha_\perp(T_N^-)$ this yields the simple $E_z H_z$-dependence of $C_1$ and $C_2$ and thus

$$\langle \alpha_{zz}(T) \rangle = \frac{1}{3}\left(\alpha_\parallel(T) + 2\alpha_\perp(T)\right)[C_1 - (C_1+C_2)Cos^3\vartheta] - (C_1+C_2)\alpha_\perp(T)Sin^2\vartheta Cos\vartheta \; . (24)$$

The dependence of $\langle \alpha_{zz}(T) \rangle$ on the annealing field product was neglected in the effective $T$=0 theory previously developed by Shtrikman and Treves **[10]**.

To further simplify expression (24) we utilize $Cos^3\vartheta \ll 1$ and $\alpha_\parallel(T_N^-) \gg -\alpha_\perp(T_N^-)$. From Eq.(22) we obtain

$$Cos\vartheta = \frac{\sqrt{-\alpha_\perp(T_N^-)}}{\sqrt{\alpha_\parallel(T_N^-) - \alpha_\perp(T_N^-)}} \approx \sqrt{-\frac{\alpha_\perp(T_N^-)}{\alpha_\parallel(T_N^-)}}$$

such that altogether $\langle \alpha_{zz}(T) \rangle$ simplifies into

$$\langle \alpha_{zz}(T) \rangle = \frac{1}{3}\Big(\alpha_\parallel(T) + 2\alpha_\perp(T)\Big)C_1 - (C_1 + C_2)\alpha_\perp(T)\sqrt{-\frac{\alpha_\parallel(T_N^-)}{\alpha_\parallel(T_N^-)}}. \qquad (25)$$

The theoretical result for $\langle \alpha_{zz}(T) \rangle$ can be plotted for various **E. H** products when substituting functional forms for $\alpha_\parallel(T)$ and $\alpha_\perp(T)$. Although theoretically motivated functional forms are not known for $\alpha_\parallel(T)$ and $\alpha_\perp(T)$, we found phenomenological expressions motivated by the fact that the ME response functions have temperature dependencies which follow the product of magnetic susceptibility and $T$-dependence of the order parameter. They read.

$$\alpha_\parallel(T) = A_\parallel\left[\left(\frac{1 + \left(\frac{T}{T_N}\right)^k}{1 + x_o^k}\right)^{\frac{1}{k}} - 1\right]\left[1 - \left(\frac{T}{T_N}\right)^{2l}\right]^{\frac{1}{l}} \qquad (26)$$

and

$$\alpha_\perp(T) = A_\perp\left[1 - \left(\frac{T}{T_N}\right)^{2l}\right]^{\frac{1}{l}} \qquad (27)$$

The expressions for $\alpha_\parallel(T)$ in equation (26) and $\alpha_\perp(T)$ in equation (27) for $x_o = 0.28$; $k = 2.8$; $l = 2.1$; $A_\parallel = \frac{32ps}{m}$; $A_\perp = -\frac{0.83ps}{m}$ reproduce the experimental $(\alpha_\parallel, \alpha_\perp)$ $vs$ $T$ curves in literature extremely well (see supplementary material for fits of Eq.(26) and Eq.(27) to experimental data from chromia single crystals) **[31]**.

The theoretical plots for $|< \alpha_{zz} >|$ $vs$ $T$ for different **E.H** products are shown in **Fig. 5**. There is a clear increase in the magnitude of $< \alpha_{zz} >$ with annealing field products as expected intuitively from the behavior of the ME response of $Cr_2O_3$ single crystals **[32]** but neglected in the existing literature of $Cr_2O_3$ powder. In addition, the inset of **Fig. 5** displays the experimental values of zero crossing temperatures for $< \alpha_{zz} >$ (circles) for increasing **E.H** products. The zero crossing temperatures increase with decreasing field product strength but level off at around 110 K. This is consistent with the subsequent theoretical discussion which reveals that isotropic ME behavior would be associated with a zero crossing of $< \alpha_{zz} >$ at 168 K. This zero crossing temperature cannot be achieve via an annealing protocol and consistently, isotropic ME response cannot be achieved via an annealing protocol for chromia powders.

**Field product dependence of the zero crossing temperature of $< \alpha_{zz} >$ $vs$. T**

The theoretical expression for the zero-crossing temperature as a function of **E.H** product is calculated from the condition $\langle \alpha_{zz}(T) \rangle = 0$. Using equation (25) for $\langle \alpha_{zz}(T) \rangle$ and the phenomenological equations (26) and (27) for the parallel and perpendicular susceptibilities one obtains the **E.H** dependent theoretical expression for the zero-crossing temperature.

$$T_{zero-crossings} = T_N \left\{ (1 + x_o^k) \left[ \frac{A_\perp}{A_\parallel} \left( 3 \left( 1 + \frac{c_2}{c_1} \right) \sqrt{\frac{-\alpha_\perp(T_N^-)}{\alpha_\parallel(T_N^-)}} + \frac{A_\parallel}{A_\perp} - 2 \right) \right]^k - 1 \right\}^{\frac{1}{k}} \quad (28)$$

The line in the inset of **Figure 5** shows the result of a least-squares fit of equation (28) to the experimental zero crossing data (circles). The nonlinear curve is fitted using $\alpha_\perp(T_N^-)$ and $\alpha_\parallel(T_N^-)$ as fitting parameters. The fit yields $\alpha_\perp(T_N^-) = -4.40715 \times 10^{-14}$ sec/m and $\alpha_\parallel(T_N^-) = 3.01693 \times 10^{-13}$ sec/m while $x_o = 0.28; k = 2.8; l = 2.1; A_\parallel = \frac{32ps}{m}; A_\perp = -\frac{0.83ps}{m}$ have been previously determined from phenomenological approximations of experimental temperature dependencies of $\alpha_\parallel$ and $\alpha_\perp$ measured in **[31]**. Using the same experimental data and comparing with the fit results $\alpha_\parallel(T_N^-)$ and $\alpha_\perp(T_N^-)$ allows to estimate $T_N^- = 306.15$ K. $T_N^-$ marks the regime shortly below the phase transition into the ordered phase. Here long range order is established on the order parameter no longer fluctuations. Therefore, as expected, $T_N^- = 306.15$ K is a value slightly below the critical temperature $T_N = 307$ K **[26, 27]**.

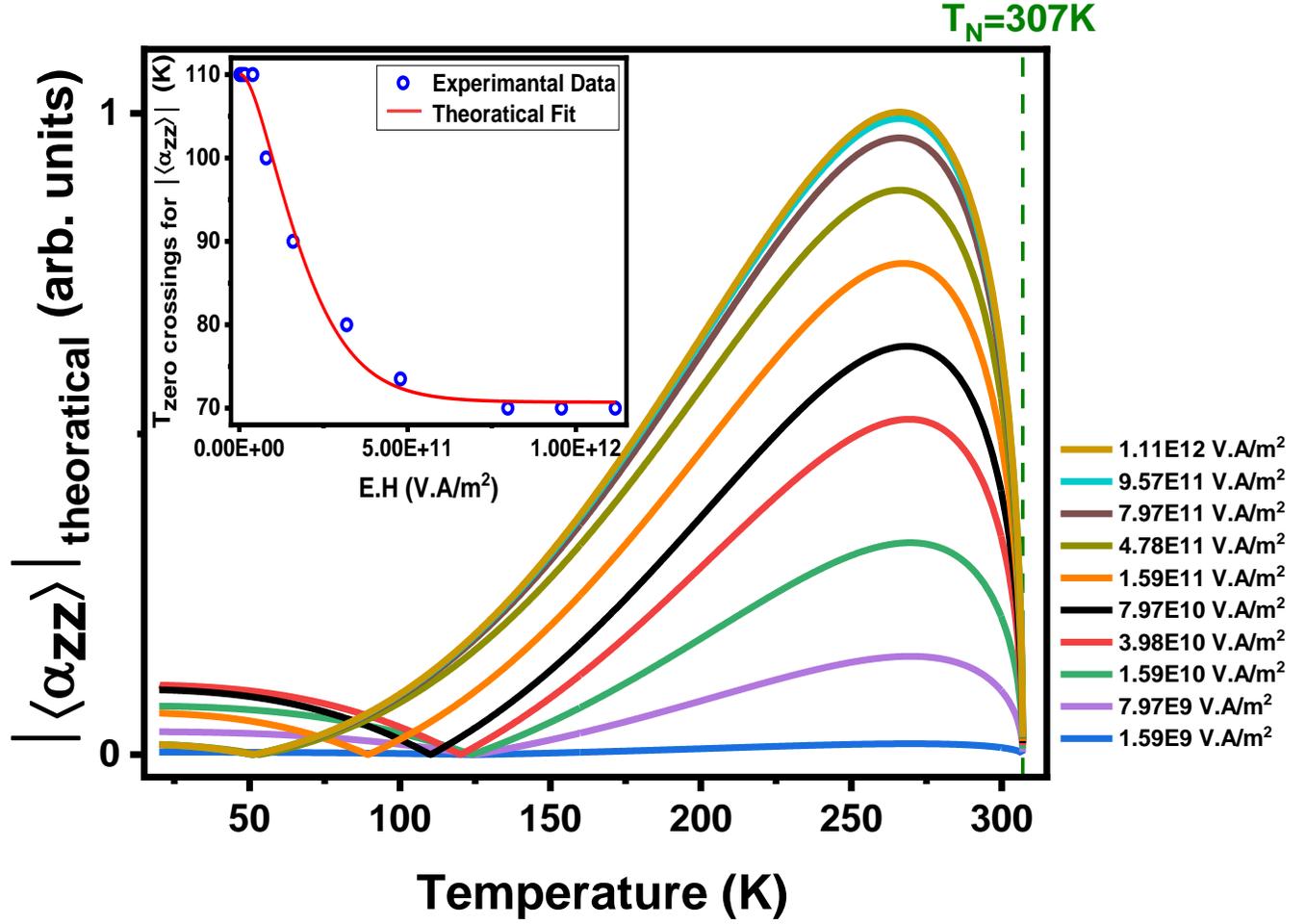

**Figure 5:** Calculated $|<\alpha_{zz}>|$ vs T (lines) for various **E.H** products. The inset shows the best fit of Eq. (28) to the experimental zero-crossing temperatures (circles) for different **E.H** products.

**Theoretical expression for the off-diagonal component $<\alpha_{zx}>$:**

The ME response $<\alpha_{zx}>$ can be obtained by first cooling the powder through $T_N$ by applying **E** field along $\hat{x}-$ axis and **H** field along $\hat{z}-$ axis and then measured by applying an AC **E** field along $\hat{x}-$ axis and detecting the magnetization response along the $z$-axis. The symmetry of the system predicts $\langle\alpha_{zx}(T)\rangle = \langle\alpha_{xz}(T)\rangle$. The expression for the off-diagonal component $\alpha_{zx}$ is calculated by using equation (18) for $i = l = z$ & $j = k = x$ as:

$$\alpha_{zx}(\theta,\phi) = \alpha_{zx}^+(T)Tanh\left[\frac{\alpha_{zx}(T_N^-)E_xH_zV}{k_BT_N}\right] \tag{29}$$

Substitute $\alpha_{zx}^+(T)$ from equation (8) in equation (29) yields

$$\alpha_{zx}(\theta,\phi) = (\alpha_{\parallel} - \alpha_{\perp})Sin\theta Cos\theta Cos\phi Tanh\left[\frac{(\alpha_{\parallel}-\alpha_{\perp})Sin\theta Cos\theta Cos\phi E_xH_zV}{k_BT_N}\right] \tag{30}$$

The average ME susceptibility response can be calculated by using the equation (19)

$$\langle\alpha_{zx}(T)\rangle = \frac{1}{4\pi}\int_0^\pi\int_0^{2\pi}\alpha_{zx}(\theta,\phi)Sin\theta d\theta\, d\phi \tag{31}$$

Using the same approximations as before, the expression for $\langle\alpha_{zx}(T)\rangle = \langle\alpha_{xz}(T)\rangle$ becomes

$$\langle\alpha_{zx}(T)\rangle = \langle\alpha_{xz}(T)\rangle \approx \frac{2}{3\pi}\left(\alpha_{\parallel}(T) - \alpha_{\perp}(T)\right)Tanh\left[\frac{2\left(\alpha_{\parallel}(T_N^-)-\alpha_{\perp}(T_N^-)\right)E_xH_zV}{\pi^2k_BT_N}\right]. \tag{32}$$

The theoretical plots for $<\alpha_{zx}>\ vs\ T$ for different **E.H** products are shown as lines in **Fig. 6**. The theoretical results also show an increase in the magnitude of $<\alpha_{zx}>$ with annealing field products as mentioned in experimental results in **Fig**. **3**.

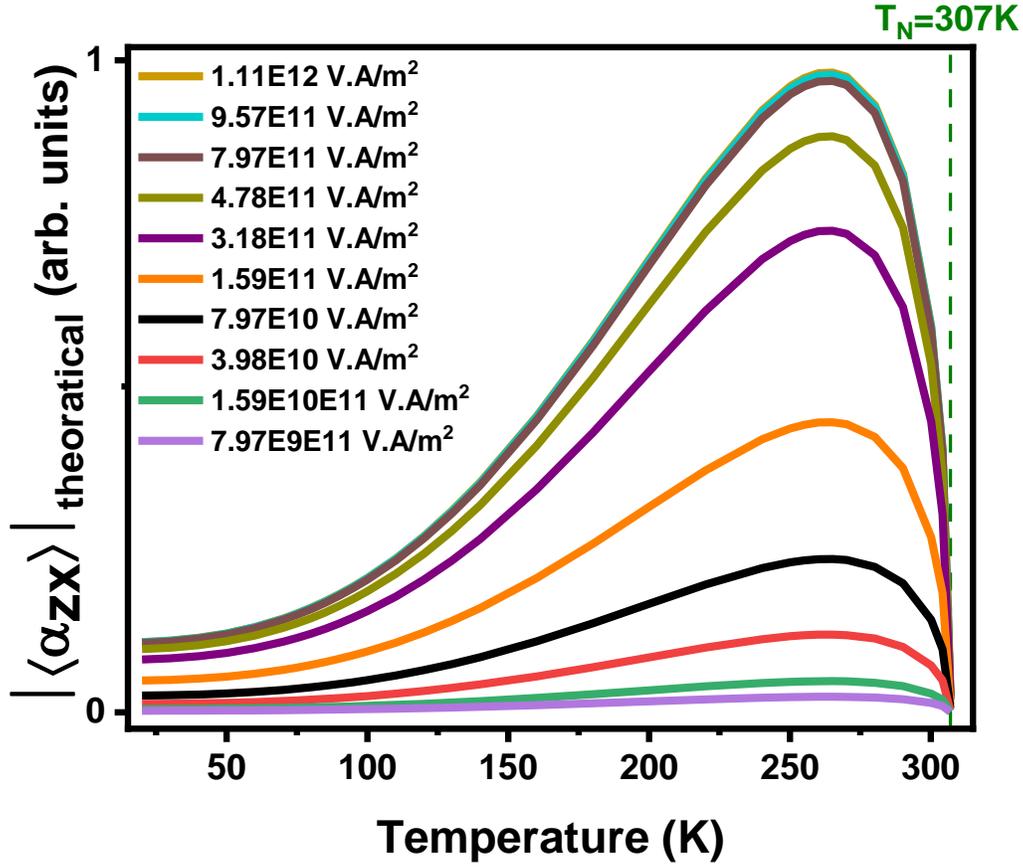

**Figure 6:** Calculated $< \alpha_{zx} >$ vs T for various **E.H** products.

**Fig. 7** shows plots of the experimentally measured off-diagonal component for various annealing field products $E_x H_z$. The experimental data are scaled by the $E_x H_z$-dependent factor in Eq.(32) according to $< \alpha_{zx} > / Tanh \left[ \frac{2 \left( \alpha_{\parallel}(T_N^-) - \alpha_{\perp}(T_N^-) \right) E_x H_z V}{\pi^2 k_B T_N} \right]$ vs T. As a result of scaling, all data points collapse within experimental scatter onto the theoretical master curve (line) $f(T) = \frac{\langle \alpha_{zx}(T) \rangle}{Tanh \left[ \frac{2 \left( \alpha_{\parallel}(T_N^-) - \alpha_{\perp}(T_N^-) \right) E_x H_z V}{\pi^2 k_B T_N} \right]} = \frac{2}{3\pi} \left( \alpha_{\parallel}(T) - \alpha_{\perp}(T) \right)$. The data collapse reveals the validity of our thermodynamic theory and highlights the hitherto suppressed fact that the ME response strongly depends on the annealing field product.

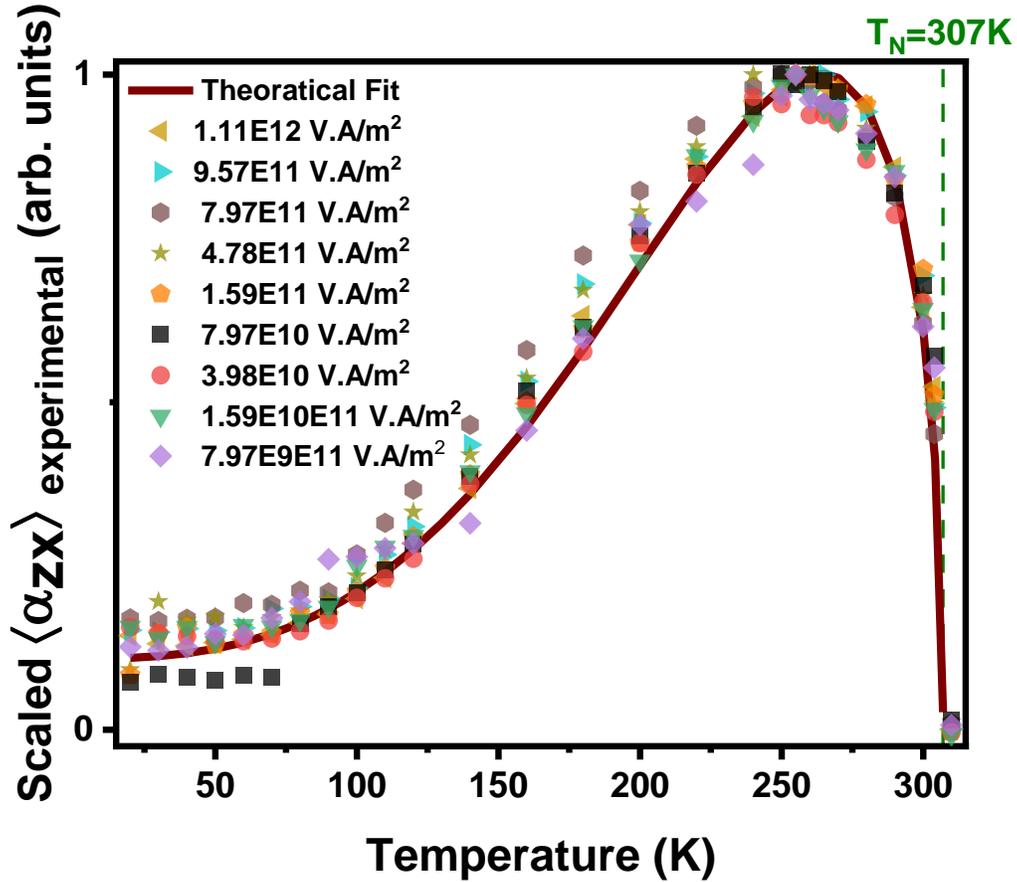

**Figure 7:** The experimentally measured $<\alpha_{zx}>$ are scaled according to $<\alpha_{zx}>/Tanh\left[\frac{2\left(\alpha_{\parallel}(T_N^-)-\alpha_{\perp}(T_N^-)\right)E_xH_zV}{\pi^2 k_B T_N}\right]$ for various annealing field products $E_xH_z$ and plotted versus $T$ (symbols). The collection of all scaled experimental data are fitted with the theoretical mast curve, $f(T)$, determine from Eq.(32) which reads $f(T)=\frac{2}{3\pi}\left(\alpha_{\parallel}(T)-\alpha_{\perp}(T)\right)$.

**Theoretical condition for pure monopole response in powder samples of ME Cr$_2$O$_3$**

The potential isotropic (monopole or axion) response is a special case given by $\langle \alpha_{xx}(T) \rangle = \langle \alpha_{yy}(T) \rangle = \langle \alpha_{zz}(T) \rangle$ and $\langle \alpha_{ij}(T) \rangle = 0$ for $i \neq j$.

To derive a hypothetical condition for isotropic response we start by calculating the average of the sum of diagonal components of (2).

$$\langle \alpha_{xx}(T) + \alpha_{yy}(T) + \alpha_{zz}(T) \rangle = \frac{1}{4\pi} \int_0^{\pi} \int_0^{2\pi} \left( \alpha_{xx} + \alpha_{yy} + \alpha_{zz} \right) Tanh \left[ \frac{\alpha_{zz} E_z H_z V}{k_B T_N} \right] Sin\theta \, d\theta \, d\phi \quad (33)$$

From equations (1) & (2) the trace of the tensor gives

$$\alpha_{xx} + \alpha_{yy} + \alpha_{zz} = \alpha_{\parallel} + 2\alpha_{\perp} \,. \quad (34)$$

Substituting equation (34) in equation (33) yields

$$\langle \alpha_{xx}(T) + \alpha_{yy}(T) + \alpha_{zz}(T) \rangle = (\alpha_{\parallel} + 2\alpha_{\perp}) \int_0^{\pi/2} Tanh \left[ \frac{\{\alpha_{\parallel}(T_N) Cos^2\theta + \alpha_{\perp}(T_N) Sin^2\theta\} E_z H_z V}{k_B T_N} \right] Sin\theta \, d\theta.$$

To evaluate the integral we use the equivalent steps applied for the derivation of $\langle \alpha_{zz}(T) \rangle$, i.e. breaking the $\theta$-integral into two parts. Utilizing the Cauchy mean value theorem and the approximations $Cos^2\theta' \approx 1/2$ for $\theta' \approx \frac{\vartheta}{2} \approx \frac{\pi}{4}$ for the integral between $[\,0, \vartheta\,]$ and for the integral between $[\,\vartheta, \pi/2\,]$ one finds $Sin^2\theta'' \approx 1$ for $\theta'' \approx \pi/2$. Together with $\alpha_{\parallel}(T_N^-) \gg -\alpha_{\perp}(T_N^-)$ this yields the simple

$$\langle \alpha_{xx}(T) \rangle + \langle \alpha_{yy}(T) \rangle + \langle \alpha_{zz}(T) \rangle = (\alpha_{\parallel} + 2\alpha_{\perp}) \left[ C_1 - (C_1 + C_2) \sqrt{\frac{-\alpha_{\perp}(T_N^-)}{\alpha_{\parallel}(T_N^-)}} \right] \quad (35)$$

The system annealed along $\hat{z}$ −axis is isotropic in the $xy$ − plane. Thus,

$$\langle \alpha_{xx} \rangle = \langle \alpha_{yy} \rangle \Rightarrow \langle \alpha_{xx} \rangle + \langle \alpha_{yy} \rangle + \langle \alpha_{zz} \rangle = 2\langle \alpha_{xx} \rangle + \langle \alpha_{zz} \rangle \quad (36)$$

Equation (36) with expression of $\langle \alpha_{zz}(T) \rangle$ from equation (25) yields the expression for $\langle \alpha_{xx}(T) \rangle = \langle \alpha_{yy}(T) \rangle$

$$\langle \alpha_{xx}(T) \rangle = \langle \alpha_{yy}(T) \rangle = \frac{1}{3}(\alpha_\parallel + 2\alpha_\perp)C_1 - \frac{1}{2}(\alpha_\parallel + \alpha_\perp)(C_1 + C_2)\sqrt{\frac{-\alpha_\perp(T_N^-)}{\alpha_\parallel(T_N^-)}} \qquad (37)$$

Comparison of $\langle \alpha_{xx}(T) \rangle$ and $\langle \alpha_{yy}(T) \rangle$ with $\langle \alpha_{zz}(T) \rangle$ from Eq. (25) reveals that isotropic response necessarily requires $\frac{1}{2}(\alpha_\parallel + \alpha_\perp)(C_1 + C_2)\sqrt{\frac{-\alpha_\perp(T_N^-)}{\alpha_\parallel(T_N^-)}} = 0$ and $(C_1 + C_2)\alpha_\perp(T)\sqrt{-\frac{\alpha_\perp(T_N^-)}{\alpha_\parallel(T_N^-)}} = 0$. Because $\alpha_\parallel \neq \alpha_\perp \, \forall \, T < T_N$ and $\alpha_\perp(T_N^-) \neq 0$ the expressions for isotropic ME response $\langle \alpha_{xx}(T) \rangle = \langle \alpha_{yy}(T) \rangle = \langle \alpha_{zz}(T) \rangle = \frac{1}{3}(\alpha_\parallel + 2\alpha_\perp)C_1$ cannot be achieved via an annealing protocol. Different experimental protocols including extremely slow cooling, rapid cooling across $T_N$ aiming at variation of $T_N^-$ were performed to minimize the $\sqrt{\frac{-\alpha_\perp(T_N^-)}{\alpha_\parallel(T_N^-)}}$ term. Although some miniscule change in $T_N^-$ and thus $\sqrt{\frac{-\alpha_\perp(T_N^-)}{\alpha_\parallel(T_N^-)}}$ can be achieved for rapid quenching toward less isotropic behavior, in general, for polycrystalline $Cr_2O_3$ powders when annealed under parallel **E** and **H** field, the ratio is always $\frac{\alpha_\perp(T_N^-)}{\alpha_\parallel(T_N^-)} < 0$ and never zero or larger zero. This contradicts the case of hypothetical isotropic ME response made in **[10]**. Hence isotropic or pure pseudoscalar ME response cannot be achieved by ME annealing a polycrystalline $Cr_2O_3$ powder. We currently explore alternative ways to start from single crystalline Chromia for the fabrication of an effective medium with pure pseudoscalar ME response. The work is in progress and will be reported elsewhere.

In conclusion we provide strong experimental evidence for the evolution of ME susceptibility in polycrystalline $Cr_2O_3$ powders with different annealing field protocols. Such dependencies are known for chromia single crystals but have not been investigated in powder samples. We determined the expressions of ME susceptibilities and the condition for isotropic (monopole, pseudoscalar or axion) behavior in polycrystalline $Cr_2O_3$ powders by incorporating Boltzmann statistics. Our findings for the polycrystalline $Cr_2O_3$ powders allow us to understand the complex variation of the temperature dependence of the diagonal and off-diagonal ME tensor elements under various ME annealing conditions. We conclude, in contrast to some claims in the literature, that pure pseudoscalar ME response cannot be achieved in $Cr_2O_3$ powders when following ME annealing protocols which start in the paramagnetic phase and end, in the presence of any electric and magnetic field configuration, in the antiferromagnetically ordered phase of the grains. This is ultimately due to the fact that $\frac{\alpha_\perp(T_N^-)}{\alpha_\parallel(T_N^-)} < 0$ . Alternative ways to utilize $Cr_2O_3$ in order to realize effective media with pure pseudoscalar ME response are currently under investigation.

**Acknowledgement:**

Financial support was provided by the NSF/EPSCoR RII Track-1: Emergent Quantum Materials and Technologies, OIA-2044049 is acknowledged and by Army Research Office through MURI W911NF-16-1- 0472. The research was performed in part in the Nebraska Nanoscale Facility: National Nanotechnology Coordinated Infrastructure and the Nebraska Center for Materials and Nanoscience, which are supported by the National Science Foundation under Award ECCS: 2025298, and the Nebraska Research Initiative.